\documentclass[11pt]{article}

\usepackage{verbatim,color,amssymb}
\usepackage{amsmath}					
\usepackage{amsthm}					
\usepackage{natbib}
\usepackage{multirow}
\usepackage{setspace}
\usepackage[mathscr]{euscript}
\usepackage{fancyhdr}
\usepackage{enumitem}
\usepackage{graphicx}
\usepackage{lineno}
\usepackage{array,booktabs}
\usepackage{url}
\usepackage{caption}
\usepackage{multibib}
\newcites{latex}{References}

\usepackage{lscape}

\usepackage{subfig}
\usepackage{tikz}
\usetikzlibrary{arrows}
\usepackage{multirow}

\newtheorem*{Proof*}{Proof}

\def\I{{\cal I}}

\def\Z{{\cal Z}}

\def\log{\hbox{log}}

\def\Bernoulli{\hbox{Bernoulli}}
\def\Beta{\hbox{Beta}}

\def\Ga{\hbox{Ga}}

\def\Normal{\hbox{Normal}}

\def\P_25_ICML{{\it Proceedings of the 25th international conference on Machine learning}}

\def\bse{\begin{eqnarray*}}
\def\ese{\end{eqnarray*}}
\def\be{\begin{eqnarray}}
\def\ee{\end{eqnarray}}
\def\bq{\begin{equation}}
\def\eq{\end{equation}}

\usepackage{verbatim,color,amssymb}

\def\bone{{\mathbf 1}}

\def\b1e{{\mathbf e}}

\def\bA{{\mathbf A}}

\def\bD{{\mathbf D}}

\def\bI{{\mathbf I}}

\def\bq{{\mathbf q}}

\def\bR{{\mathbf R}}

\def\bx{{\mathbf x}}
\def\bX{{\mathbf X}}
\def\by{{\mathbf y}}

\def\bz{{\mathbf z}}

\newcommand{\bmu}{\mbox{\boldmath $\mu$}}

\newcommand{\btheta}{\mbox{\boldmath $\theta$}}
\newcommand{\bbeta}{\mbox{\boldmath $\beta$}}
\newcommand{\bgamma}{\mbox{\boldmath $\gamma$}}

\newcommand{\bGamma}{\mbox{\boldmath $\Gamma$}}
\newcommand{\btau}{\mbox{\boldmath $\tau$}}

\renewcommand\footnoterule{\kern-3pt \hrule \textwidth 2in \kern 2.6pt}

\def\boxit#1{\vbox{\hrule\hbox{\vrule\kern6pt \vbox{\kern6pt \textcolor{blue}{#1}\kern6pt}\kern6pt\vrule}\hrule}}

\def\authorfootnote#1{{\let\thefootnote\relax\footnotetext{#1}}}
\allowdisplaybreaks
\usepackage{xr}
\makeatletter

\newcommand*{\addFileDependency}[1]{% argument=file name and extension
\typeout{(#1)}% latexmk will find this if $recorder=0
\@addtofilelist{#1}
\IfFileExists{#1}{}{\typeout{No file #1.}}
}\makeatother

\newcommand*{\myexternaldocument}[1]{%
\externaldocument{#1}%
\addFileDependency{#1.tex}%
\addFileDependency{#1.aux}%
}
\myexternaldocument{supplementary}

\begin{document}

% Title of paper
\title{Individualized Multi-Treatment Response Curves Estimation using RBF-net with Shared Neurons}

% List of authors, with corresponding author marked by asterisk
\author{PETER CHANG, ARKAPRAVA ROY\\
% Author addresses
\textit{Department of Biostatistics,
University of Florida,
%College of Public Health and Health Professions,
%Gainesville, Florida,
United States}
\\
% E-mail address for correspondence
{pchang27@ufl.edu, arkaprava.roy@ufl.edu}
}

% Running headers of paper:
\markboth%
% First field is the short list of authors
{CHANG, ROY}
% Second field is the short title of the paper
{Multi-treatment RBF-net}

\maketitle

% Add a footnote for the corresponding author if one has been
% identified in the author list
% \footnotetext{To whom correspondence should be addressed.}

\begin{abstract}
{Heterogeneous treatment effect estimation is an important problem in precision medicine. 
Specific interests lie in identifying the differential effect of different treatments based on some external covariates.
We propose a novel non-parametric treatment effect estimation method in a multi-treatment setting.
Our non-parametric modeling of the response curves relies on radial basis function (RBF)-nets with shared hidden neurons. Our model thus facilitates modeling commonality among the treatment outcomes.
The estimation and inference schemes are developed under a Bayesian framework using thresholded best linear projections and
implemented via an efficient Markov chain Monte Carlo algorithm, 
appropriately accommodating uncertainty in all aspects of the analysis. 
The numerical performance of the method is demonstrated through simulation experiments. 
Applying our proposed method to MIMIC data, we obtain several interesting findings related to the impact of different treatment strategies on the length of ICU stay and 12-hour SOFA score for sepsis patients who are home-discharged.}

\end{abstract}

\newpage

\section{Introduction}
In many clinical settings, such as intensive care units (ICUs), patients often receive multiple treatments simultaneously. For instance, in the MIMIC database, septic patients frequently receive both mechanical ventilation and vasopressors, either separately or simultaneously \citep{peine2021development, zhu2021machine, xu2023timing, he2023effect}. Similarly, in clinical trials, treatment arms often share a common drug across different interventions \citep{james2016addition, ursprung2021wire}, reflecting potential commonalities in treatment effects. Recognizing that treatments may share underlying mechanisms, which influence the outcomes \citep{jones2003comparison, white2011effects}, it becomes important to adequately account for these shared characteristics in the model. This motivates the development of the proposed model to estimate heterogeneous treatment effects with shared characteristics.

Estimating heterogeneous treatment effects from observational data is crucial for determining individualized causal effects \citep{wendling2018comparing} and learning optimal personalized treatment assignments \citep{rekkas2020predictive, zhang2012robust, %luedtke2016statistical, 
jiang2017estimation, cui2021semiparametric, li2023optimal}. However, accurate estimation requires reasonable representations from each treatment subgroup.
The advent of large-scale health outcome data, such as electronic health records (EHRs), has significantly optimized the design of several novel statistical methods, primarily tailored for binary treatment scenarios \citep{wendling2018comparing, cheng2020estimating}, with some accommodating multiple treatment settings \citep{brown2020novel, chalkou2021two}.
While many approaches focus on estimating population average treatment effects (ATEs) \citep{van2006tmle, chernozhukov2018double, mccaffrey2013tutorial}, newer methods aim to estimate conditional average treatment effects (CATEs) \citep{taddy2016nonparametric,wager2018causalforest,kunzel2019metalearners}. 

In a binary setting, the treatment effect $\tau(\bx)$ can be estimated as a function of the predictor vector $\bx$ with a mean-zero additive error model. Here, the continuous treatment outcome $y_i$ is expressed as $y_i=f(\bx_i, z_i) + \epsilon_i$, where $z_i$ is a treatment indicator for the $i$-$th$ subject. Under this model, $\tau(\bx_i)=f(\bx_i, 1)-f(\bx_i, 0)$. Another possibility $f(\bx_i, z_i)=\mu(\bx_i)+\tau(\bx_i)z_i$ \citep{hahn2020bayesian} can also be used for a direct estimation of the treatment effect. Alternatively, one may model each treatment group
separately with two functions $f_0(\bx_i)$  and $f_1(\bx_i)$ and define $\tau(\bx_i)=f_1(\bx_i)-f_0(\bx_i)$. These two approaches are defined as S-learner and T-learner, respectively \citep{kunzel2019metalearners}. 
%In addition, \cite{kunzel2019metalearners} proposes the X-learner approach for CATE estimation. 
On Bayesian causal inference, \cite{li2023bayesian} provides a critical review.
%Another approach called R-leaning is recently proposed by \cite{nie2021quasi}, which is implemented in the {\tt multi\_arm\_causal\_forest} function of R package {\tt grf} \citep{grfR}. 
In a multi-treatment setting, one may either develop a model with the treatment indicator as a predictor as in S-learning or set treatment-specific functions as in  T-learning. 
However, traditional T-learners do not work very well when a common behavior is expected in two response functions, as shown in \cite{kunzel2019metalearners} for a two-treatment design.  

To address these limitations, we propose a radial basis function-based modeling approach that accommodates shared characteristics between outcomes and captures both shared and treatment-specific features. Our model allows the recognition of treatments sharing common underlying pathways or characteristics, thus improving the accuracy of treatment effect estimates. Here, we tackle a generic problem of heterogeneous treatment effect or CATE estimation in a multi-treatment setting, where the treatment responses may share some commonalities. 
The proposed model can be applied in estimating outcome functions, particularly when they are expected to share common characteristics.

We further propose a thresholding-based inference procedure relying on the best linear projections from \cite{cui2023estimating}. Specifically, we first modify the approach from \cite{cui2023estimating} for inference using the posterior samples and then propose a variable importance estimation procedure by combining a thresholding step.
The estimated importance values are shown to work reasonably well in simulation and are later used for inference in our MIMIC application.

The remainder of this article is organized as follows. The subsequent section introduces our proposed modeling framework. Section~\ref{sec:bayes} discusses our Bayesian inference scheme detailing the prior specification, likelihood computation, detailed posterior computation steps, and linear projection-based inference scheme. Subsequently, in Section~\ref{sec:simu}, we analyze and compare the effectiveness of our proposed method against other competing approaches under various simulation settings. Finally, we apply our proposed method to study the effect of different treatments on patients, who suffered from acute sepsis and were discharged to home at the end of their ICU visit using MIMIC data in Section~\ref{sec:real}, and Section~\ref{sec:discussion} is dedicated to discussing the overall conclusion and potential extensions.

\section{Model} \label{sec:model}
\begin{figure}
    \centering
    \includegraphics[scale=1]{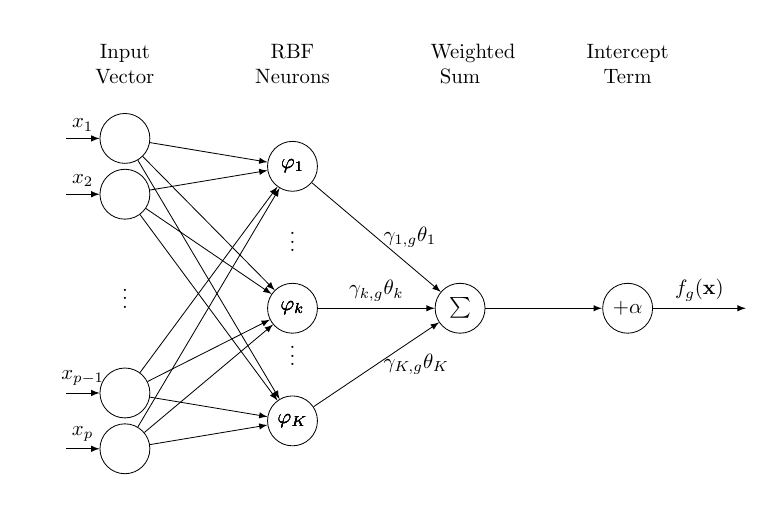}
    \caption{A schematic representation of our proposed RBF-net where $\varphi_k(x)= \exp\{-(\bx_{i}-\bmu_k)^T(\bx_i-\bmu_k)/b_k^2\}$. The activation functions are varying at each hidden unit due to the different centers $\bmu_k$. The $\gamma_{k,g}$'s are binary and control the inclusion of the neurons.}
    \label{fig:galaxy}
\end{figure}

Our proposed model falls under the class of T-learner type models.
We consider $G$ treatments and assume that there are $P$ covariates.
Let $y_{i}$ be a continuous treatment outcome for the $i$-$th$ patient with subject-specific $P$-dimensional covariates specified by $\bx_{i}$.  
Then, our model for $y_{i}$ when treated under the $g$-$th$ treatment is,
%We propose a modeling framework for estimating treatment effects when there are multiple treatments. The data is generally available as (Outcome, Covariates, Treatment) = $(y_i,\bx_i,g)$. For simplicity, we first consider the case with $y_i$ being a continuous outcome. Let us assume that there are $G$ many available treatments.
\begin{align}
    y_{i}=&f_{g}(\bx_{i}) + \epsilon_{i},\\%\nonumber\\
    \epsilon_i\sim&\Normal(0,\sigma^2),\\%\nonumber\\
    f_{g}(\bx_{i})=&\alpha+\sum_{k=1}^K\gamma_{k,g}\theta_{k}\exp\{-(\bx_{i}-\bmu_k)^T(\bx_i-\bmu_k)/b_k^2\}. \label{eq:model}
%\gamma_{1,d},\ldots,\gamma_{K,d}\sim&\Bernoulli(p_d), \quad p_d\sim\Unif(0,1)
\end{align}
Here, $f_1(\cdot),\ldots,f_G(\cdot)$ denote the treatment-specific outcome functions modeled using a common set of radial basis function (RBF) neurons. Each function has a treatment-specific parameter $\bgamma_{g}=(\gamma_{k,g})_{1\leq k\leq K}$ in its coefficients, where $\gamma_{k,g}\in\{0,1\}$, and $\bmu_k, \theta_k\in\mathbb{R}$. %We assume that .
%These indicators help to drop out some of the components (this is a popular architecture used to fit neural nets).
%Specifically, for treatment group 1 i.e. for subjects with $g=1$, we have the treatment effect as $f_{1}(\bx)=\sum_{k=1}^K\theta_{k}\gamma_{k,1}\exp\{-(\bx-\bmu_k)^T(\bx_i-\bmu_k)/\sigma_k^2\}$ which is popularly known as the RBF-net. 
We then construct the parameter $\bGamma=(\!(\gamma_{k,g})\!)_{1\leq k\leq K, 1\leq g\leq G}$ as a $K\times G$ dimensional binary matrix. This matrix plays a crucial role in our proposed model, controlling the set of RBF bases to be used in specifying a treatment-specific outcome function.
Here, $\gamma_{k,g}$ is 1 whenever the $k$-$th$ RBF basis is included in modeling $f_g(\bx)$. A schematic representation of the proposed mode is in Figure~\ref{fig:galaxy}. Note that the flexibility of the proposed model automatically extends to the cases with minimal to no sharing among the $f_g$'s. 
%\subsection{Inference}

 %In our modeling framework, although $\bGamma$ controls shared and treatment-specific RBF bases, we are not specifically interested in identifying similarities in the response functions using this parameter. 
%, which are often inefficient in estimating the treatment-specific functions sharing a common behavior, as pointed out in \cite{kunzel2019metalearners}.
The role of $\bGamma$ is to address the inefficiency mentioned previously about T-learner type models. Though $\bGamma$ determines the shared and treatment-specific RBF bases, its primary role is to improve the estimation of outcome functions that may share common characteristics, rather than explicitly identifying similarities and differences between responses.

In our proposed model, the conditional treatment effect of $g$ over $g'$ for an array of covariates $\bx$ is identified as $\tau_{gg'}=f_g(\bx)-f_{g'}(\bx)$ under assumptions that 1) given $\bx$, treatment assignments are random and do not depend on the outcome, 2) the overlap condition $1>P(z_i=g\mid\bx_i)>0$ holds for all $i$, where $z_i$ is the treatment assignment variable such that $z_i=g$ implies $i$-$th$ subject is treated under $g$-$th$ treatment and 3) standard stable unit treatment value assumption (SUTVA) is maintained requiring no interference that one subject’s potential outcomes are not affected by other subjects’ treatment assignments.

\section{Bayesian inference}\label{sec:bayes}

We pursue a Bayesian approach to carry out the inference of the proposed RBF-net model with shared neurons, aiming to automatically quantify estimation uncertainty.
In our model, $\{\Gamma, \bmu, \alpha, \btheta,\sigma, K\}$ is the complete set of parameters. Here,  $\alpha \in \mathbb{R}$, $\sigma \in \mathbb{R}$, and $K \in \mathbb{R}$. As stated in Section~\ref{sec:model}, $\bGamma=(\!(\gamma_{k,g})\!)_{1\leq k\leq K, 1\leq g\leq G}$ is a $K\times G$ dimensional binary matrix. The RBF-centers $\bmu = \{\bmu_1, ..., \bmu_K \}$ is $P \times K$ dimensional matrix, where $\bmu_k = (\!(\mu_{p,k})\!)_{1\leq p\leq P}$. $\alpha$ is the intercept, $\btheta = \{\theta_1, \cdots, \theta_k\}$ is the $K$-length vector of coefficients, $\sigma$ is the error standard deviation, and finally $K$ is the total number of RBF-bases. Although $K$ is unknown, we first specify the priors assuming a given value of $K$. In Section~\ref{sec:specifyK}, our specification procedure for $K$ is discussed following an empirical Bayes-type strategy.

\subsection{Prior specification} \label{sec:prior}

To proceed with our Bayesian analysis, we put prior distributions on the parameters. We assume independent priors for $\bGamma, \bmu, \alpha, \btheta, \sigma$. %Most of the priors are also weakly informative unless specfied. 
They are described below in detail.

\begin{enumerate}
     \item The basis inclusion indicators $(\bGamma)$: The indicators are independent and follow Bernoulli distribution as $\gamma_{k,g} \sim  \text{Bernoulli}(p_g)$ where $p_g\sim \Beta(c,d)$. %=& p_g^{\gamma_{k,g}} (1-p_g)^{1-\gamma_{k,g}}
\item The RBF-centers $(\bmu)$: We put a multivariate normal distribution as the prior for $\bmu$,
$\mu_{p,k} \sim  \Normal(0, \sigma_{\mu}^2)$.  %=& \exp \left(- \frac{\mu_{p,k}^2 }{2 \sigma_{\mu}^2} \right)\\
\item The intercept and coefficients $(\alpha, \btheta)$: We let both $\alpha, \theta_k  \sim  \Normal(0, \sigma^2_{d})$, with a pre-specified $\sigma_d$. %$\sigma_d\sim C^{+}(0,A_d)$. %, where $C^{+}$ denotes the half-Cauchy distribution. %=& \exp \left(- \frac{\theta_k^2}{2 \sigma_d} \right); \sigma_d = k^{-1} \\
\item The error variance ($\sigma$): We specify a half-Cauchy prior for $\sigma$ as $\sigma \sim C^{+}(0,\hat{\sigma})$. 
    \end{enumerate}

%For the RBF-centers, setting the prior mean to be 0 is reasonable when centering and scaling both the outcome and predictors. 
%In the case of the variance for $\sigma_{\mu}$, we set $\sigma_{\mu}=1$ as we normalize the predictors. 
The choice of the half-Cauchy distribution as prior for $\sigma$ is motivated by \cite{linero2018bayesian}. %For the error variance, $\hat{\sigma}$ is an estimate of $\sigma$ from the data \citep{chipman2010bart}. 
We set $\hat{\sigma}$ to be the error standard deviation from the linear regression model, where the outcome is regressed on the predictors and the treatment covariate as a factor. Further details on the hyperparameters $c$, $d$, $\sigma_{\mu}$, and $\sigma_d$ is provided in Section \ref{sec:hyperparameters}.

\subsection{Likelihood function of data}

In this section, we formalize the posterior likelihood function for our Bayesian inference model, considering both the observed data and the relevant parameters. 
%In the univariate case, the expression for our joint posterior probability of our model can be written as : $$p(\gamma_{k,g}, \mu_{p,k}, \theta_k, \mu, \sigma | y_i) \propto p(y_i | \gamma_{k,g}, \mu_{p,k}, \theta_k, \mu, \sigma) \times p(\gamma_{k,g}) \times p(\mu_{p,k}) \times p(\theta_k) \times p(\mu) \times p(\sigma)$$
The likelihood of our proposed model is \begin{equation} p(\by | \bGamma, \bmu, \btheta, \alpha, \sigma) \propto \prod_i^N \prod_g^G \exp \left(- \frac{(y_{i}- f_{g}(\bx_{i}))^2 }{2 \sigma^2} \right),\end{equation} %\left(\frac{1}{\sigma} \right)
and with the prior specified in Section \ref{sec:prior} of the main paper, we can formally write each prior as

\begin{align}
p(\bGamma) \propto& \prod_g^G \prod_k^K p_g^{\gamma_{k,g}} (1-p_g)^{1-\gamma_{k,g}}\\
p(\bmu) \propto&  \prod_p^P \prod_k^K \exp \left(- \frac{\mu_{p,k}^2 }{2 \sigma^2_\mu} \right)\\
p(\btheta) \propto& \prod_{k}^{K} \exp\left(-\frac{\theta_k^2}{2\sigma_d^2}\right) \\%\times \exp\left(-\frac{\alpha^2}{2\sigma_d^2}\right) \times \left(\sigma_d \left[1 + \left(\frac{\sigma_d}{A_d}\right)^2\right]\right)^{-1}\\
p(\alpha) \propto& \exp\left(-\frac{\alpha^2}{2\sigma_d^2}\right) \\
p(\sigma) \propto&  \left(\hat{\sigma}\left[1+\left(\frac{\sigma}{\hat{\sigma}}\right)^2\right]\right)^{-1}
\end{align}

The full proportional posterior is then given by:

\begin{align*}
  p(\bGamma, \bmu, \btheta, \alpha, \sigma | \by) \propto
  & \prod_i^N \prod_g^G \prod_p^P \prod_k^K \left(\frac{1}{\sigma} \right) \exp \left(- \frac{(y_{i}- f_{g}(\bx_{i}))^2 \sigma^2}{2 \sigma^2} \right) \times  p_g^{\gamma_{k,g}} (1-p_g)^{1-\gamma_{k,g}} \\
  &  \times \exp \left(- \frac{\mu_{p,k}^2 }{2 \sigma^2_\mu} \right) \times \exp \left(- \frac{ \theta_k^2}{2\sigma^2_d} \right) \times \exp\left(-\frac{\alpha^2}{2\sigma_d^2}\right)\\
  & \times \left(\hat{\sigma}\left[1+\left(\frac{\sigma}{\hat{\sigma}}\right)^2\right]\right)^{-1} 
\end{align*}

In the next subsections, we discuss our posterior sampling algorithms, the specification of the prior hyperparameters, and other related computational details.

\subsection{Posterior sampling}

We use superscript $(t)$ to represent the corresponding posterior sample at the $t$-$th$ iteration while describing the MH steps in detail below. Our sampler iterates between the following steps:

\begin{enumerate}
    
    \item \textbf{Updating $\alpha$, $\btheta$:} 
    The parameter $\alpha$ is updated by sampling from a normal distribution, $\alpha \sim \Normal(\alpha_0 \sigma_0, \sqrt{\sigma_0})$. Here, $\alpha_0$ is defined as $\sum_i(y_i - f_{g}(\bx_{i}))/\sigma^{2}$, and $\sigma_0$ as $(\sigma_d^{-2} + N/\sigma^{2})^{-1}$.
    
    The full conditional of $\btheta$ is $\Normal(\Tilde{\btheta}, \bD)$, where $\Tilde{\btheta} = \bD \left(\sigma_d^{-2} \btheta_0 + \bX^{*T} \bR /\sigma^{2}  \right)$ and $\bD = (\sigma_d^{-2}\bI_K + \bX^{*T}\bX^{*}/\sigma^2)^{-1}$.
    Here we define $\btheta_0 = (0, \cdots, 0)^T$ and $\bX^{*}$ as a $N \times K$ matrix, where the $i,k$-$th$ entry is $X^{*}_{i,k} = \gamma_{k,g} \exp\{-(\bx_{i}-\bmu_k)^T(\bx_i-\bmu_k)/b_k^2\}$, in which $g$ is the corresponding treatment group for the $i$-$th$ patient. We let the vector $\bR = (R_1, \cdots, R_n)$ be the residuals $R_i = y_i - \alpha$. 

    In cases of large values of $K$, the above-mentioned sampling method for $\btheta$ may not be efficient, particualrly when updating $\btheta$ from the full conditional where the $k$-$th$ column of $\bX^{*}$ contains only 0. This condition implies that the corresponding $k$-$th$ row of $\bGamma$ is entirely 0. However, there is a strategic approach to mitigate this potential issue. When the $k$-$th$ row of $\bGamma = 0$, the full conditional of $\btheta$ simplifies to the prior distribution. We can then use a two-part sampling strategy for $\btheta$. We create a set of indices $\Z$ such that the $z$-$th$ row in which $\bGamma$ contains only 0 for all $z \in \Z$, and sample $\theta_z$ from the prior distribution $\Normal(0, \sigma^2_{d})$. Let $\Z^c$ be the complement set of indices. We sample $\btheta_{\Z^c}$ from the conditional $\Normal(\Tilde{\btheta}_{\Z^c}, \bD_{\Z^c})$. In this case $\Tilde{\btheta}_{\Z^c} = \bD_{\Z^c} \left(\sigma_d^{-2} \btheta_{0,\Z^c} + \bX_{\Z^c}^{*T} \bR /\sigma^{2}  \right)$, where $\bX_{\Z^c}^{*T}$ is the submatrix from $\bX^{*}$ with columns from $\Z^c$, and $\bD_{\Z^c} = (\sigma_d^{-2}\bI_{|\Z^c|} + \bX_{\Z^c}^{*T}\bX_{\Z^c}^{*}/\sigma^2)^{-1}$, where $|\Z^c|$ denotes the number of entries $\Z^c$. This approach reduces computation time for large values of $K$.

    % In cases of large values of $K$, the above-mentioned sampling method for $\btheta$ may not be efficient.  This is evident when updating $\btheta$ from the full conditional despite when the $k$-$th$ row of $\bGamma = 0$. However, there is a strategic approach to mitigate this potential issue. When the $k$-$th$ row of $\bGamma = 0$, the full conditional of $\btheta$ simplifies to the prior distribution. We can then use a two-part sampling strategy for $\btheta$: for any non-zero columns in $\bX^{*}$, we sample $\btheta$ from the full conditional. In the other columns of $\bX^{*}$, we sample $\btheta$ from the prior. 
    
     %{\color{red} If $\btheta_0 = (0, \cdots, 0)^T$, then $\bD \left(\b\sigma_d^{-1} \btheta_0 + \sigma^{-2} \bX^{*T} \bR \right)$ can be written as $\bD \left(\sigma^{-2} \bX^{*T} \bR \right)$?}
    \item \textbf{Updating $\bGamma$:} Here, we consider full conditional Gibbs updates, where we update $\gamma_{k,g}$ in random order. We update $\gamma_{k,g}$ by sampling from a Bernoulli distribution, $\gamma_{k,g} \sim \Bernoulli(\Tilde{p}_g)$, where $\Tilde{p}_g = \frac{{p}_g l^1_g}{{p}_g l^1_g + (1-{p}_g)l^0_g}$. Here $l^1_g$ is the model likelihood when $\gamma_{k,g} = 1$ and likewise, $l^0_g$ is the model likelihood when $\gamma_{k,g} = 0$.

    \item \textbf{Updating ${p}_g$}: We sample ${p}_g \sim \Beta(c + \sum_g \gamma_{k,g}, d + K - \sum_g \gamma_{k,g})$. However, as discussed in Section \ref{sec:specifyK} of the main paper, some of the indices are kept fixed for the first 1000 iterations of the MCMC and thus, the above posterior distribution is marginally adjusted by excluding those fixed indices.

We now describe the updates for ($\sigma$ and $\bmu_k$), %$\sigma_d$, and $\bmu_k$)
employing the Metropolis-Hastings (MH) sampling algorithm. For $\sigma$, we use the MH step as proposed in  \cite{linero2018bayesian}, while for $\bmu_k$ a gradient-based Langevin Monte Carlo (LMC) sampling algorithm is adopted. This approach is beneficial for complex hierarchical Bayesian model. We use superscript $(t)$ to represent the corresponding posterior sample at the $t$-$th$ iteration while describing the MH steps in detail below.

\end{enumerate}
  
\begin{enumerate}%[resume]
    \item \textbf{Updating $\sigma$}:  We implement the updating technique from \cite{linero2018bayesian}. When imposing a flat prior for $\sigma^{-2}$, we can show that the conditional distribution for $\sigma^{-2}$ is $\sigma^{-2} \sim \Ga(1 + N/2, \sum_g \sum_{i\in g} (R_i -  f_{g}(\bx_{i}))^2)$, where Ga stands for the gamma distribution. Under a flat prior, we can let this full conditional be our proposal distribution. This idea enables there to be a simplification in the acceptance probability, in which the acceptance probability conveniently becomes the ratio of priors. However, in order to integrate the proposal distribution in the standard MH algorithm, we need to adjust the acceptance probability with the Jacobian transformation. Thus, we can accept the proposal with probability \begin{equation}A\left(\sigma^{(t)}  \rightarrow \sigma^{\prime}\right)=\frac{C^{+}\left(\sigma^{\prime} \mid 0, \widehat{\sigma}\right) \sigma^{\prime 3}}{C^{+}\left(\sigma^{(t)}  \mid 0, \widehat{\sigma}\right)\sigma^{(t)3} } \wedge 1 \end{equation}

     \item \textbf{Updating $\bmu_k$:} To update $\bmu_k$, we use a gradient based sampler. The conditional posterior log-likelihood and its derivative with respect to $\mu_{p,k}$ are needed this process. The conditional posterior log-likelihood $\log P(\bmu_{k} \mid \by, \bGamma, \btheta, \alpha, \sigma)$ is %given by 
     $$-\frac{n}{2} \log (2 \pi \sigma^2) - \frac{1}{2 \sigma^2} \sum_{i=1}^n (R_i- f_{g}(\bx_{i}))^2 - \sum_{p=1}^P \frac{\mu_{p,k}^2}{2 \sigma^2_\mu}.$$
    % \begin{equation}\log P(\bmu_{k} \mid \by, \bGamma, \btheta, \alpha, \sigma) = -\frac{n}{2} \log (2 \pi \sigma^2) - \frac{1}{2 \sigma^2} \sum_{i=1}^n (R_i- f_{g}(\bx_{i}))^2 - \sum_{p=1}^P \frac{\mu_{p,k}^2}{2 \sigma^2_\mu}.\end{equation} 
     The $p$-$th$ entry of corresponding derivative $\nabla \log P(\bmu_{k} \mid \by, \bGamma, \btheta, \alpha, \sigma)$, with respect to $\mu_{p,k}$, is given by $$\frac{2}{\sigma^2} \theta_k \gamma_{k, g} \sum_{i=1}^n  \exp\left(-\sqrt{\frac{\sum_{p=1}^P (x_{i,p} - \mu_{p,k})^2}{b_k^2}}\right) \frac{\sum_{p=1}^P (x_{i,p} - \mu_{p,k})^2}{b_k^2} \left(\frac{x_{i,p} - \mu_{p,k}}{b_k^2}\right)  - \frac{\mu_{p,k}}{\sigma^2_\mu}.$$
     %\begin{equation}\frac{2}{\sigma^2} \theta_k \gamma_{k, g} \sum_{i=1}^n  \exp\left(-\sqrt{\frac{\sum_{p=1}^P (x_{i,p} - \mu_{p,k})^2}{b_k^2}}\right) \frac{\sum_{p=1}^P (x_{i,p} - \mu_{p,k})^2}{b_k^2} \left(\frac{x_{i,p} - \mu_{p,k}}{b_k^2}\right)  - \frac{\mu_{p,k}}{\sigma^2_\mu}.\end{equation} 

    With gradient sampling, our proposed value now depends on the derivative, %\log P(\bmu_{k})$, 
    so that $\bmu_{k}^\prime=\bmu_{k}^{(t)}+\frac{\epsilon}{2} \nabla \log P(\bmu_{k} \mid \by, \bGamma, \btheta, \alpha, \sigma) \mid_{\bmu_{k}=\bmu_{k}^{(t)}} %\log P(\bmu_{k})
    +\delta$, where $\delta \sim \operatorname{Normal}(0, \epsilon)$. 
    % Our proposal distribution uses the expectation and variance of $\bmu_{k}^\prime \mid \bmu_{k}^{(t)}$, where 
    % \begin{align*}
    %     \mathrm{E}(\bmu_{k}^\prime \mid \bmu_{k}^{(t)}) =& \bmu_{k}^{(t)}+\frac{\epsilon}{2} \nabla \log P(\bmu_{k} \mid \by, \bGamma, \btheta, \alpha, \sigma) \mid_{\bmu_{k}=\bmu_{k}^{(t)}} \\
    %     \mathrm{Var}(\bmu_{k}^\prime \mid \bmu_{k}^{(t)}) =& \epsilon
    % \end{align*}
    Thus, the transition probability is \begin{equation} q\left(\bmu_{k}^\prime \mid \bmu_{k} \right) \propto \exp \left(-\frac{1}{2 \epsilon}\left\|\bmu_{k}^\prime-\bmu_{k}-\frac{\epsilon}{2} \nabla \log P(\bmu_{k} \mid \by, \bGamma, \btheta, \alpha, \sigma) \right\|_2^2\right),\end{equation}  and  then the acceptance probability for our gradient sampler for a proposed value $\bmu^{\prime}$ is \begin{equation}A\left(\bmu_{k}^{(t)} \rightarrow \bmu_{k}^\prime\right)=\frac{P(\bmu_{k}^\prime \mid \by, \bGamma, \btheta, \alpha, \sigma) q(\bmu_{k}^{(t)}  \mid \bmu_{k}^\prime)}{P(\bmu_{k}^{(t)}  \mid \by, \bGamma, \btheta, \alpha, \sigma) q(\bmu_{k}^\prime \mid \bmu_{k}^{(t)} )} \wedge 1.\end{equation}  The proposal values for $\mu_{p,k}$ are adjusted to achieve a pre-specified acceptance rate, tuning $\epsilon$ every 200 iterations to maintain an acceptance rate between $0.45$ and $0.7$.
   % Proposal distribution: Normal$(\bmu_{k}^{(t)}+\frac{\epsilon}{2} \nabla \log P(\bmu_{k})\mid_{\bmu_{k}=\bmu_{k}^{(t)}}, \epsilon)$$

\end{enumerate}

\subsection{Specification of $K$}
\label{sec:specifyK}

Our approach here relies on the empirical Bayes-type scheme, discussed in \cite{roy2023nonparametric}. We first create an index list to group the response and predictors associated with the $g$-$th$ treatment. For each treatment group, we utilize Gaussian RBF kernel-based relevance vector machine (RVM) within the {\tt rvm} function within the R package {\tt kernlab} \citep{kernlab}, which provides the group-specific number of relevance vectors $v_g$. We then set $K = G \times \max_g v_g$, combining all the outputs of {\tt rvm}. 

For initializing $\bGamma$, we rely on both $v_g$ and $K$. We first specify $G$ disjoint sets, denoted as $\{\I_1,\ldots,\I_G\}$. We have $\I_g\cap \I_{g'}=\emptyset$ for all $g\neq g'$ and $\cup_{g=1}^G \I_g \subset \{1,2,\ldots,K\}$. These are a small set of indices such that $\gamma_{k,g}$ is fixed at 1 for all $k\in \I_{g}$ at the initial stage of the MCMC. The set $\I_g$ is constructed as $\I_g = \{ k + \sum_{t=1}^{g-1} v_t : k \in \{1, 2, \ldots, v_g\} \}$. For the first 1000 iterations, we only sample $\gamma_{k,g}$ for $k \notin \I_{g}$, and then for all $k$ afterwards.

The initialization of the RBF centers $\bmu$ relies on the entropy-weighted k-means (EWKM) clustering algorithm, which incorporates an entropy-based weighting scheme on traditional k-means clustering.
%The EWKM algorithm is based on traditional k-means clustering by incorporating an entropy-based weighting scheme \citep{ewkm2007lpj}. 
We implement the algorithm through the {\tt ewkm} function from the {\tt wskm} package in R \citep{wskm2014hz}, where we set $K$ as the target number of centers. The function returns the centroids of the clusters, computed as the weighted mean position of all points in each cluster. To ensure stability, we add a small amount of random noise (from a normal distribution) to the centroids, as some initial centers are identical in the {\tt ewkm} output. The final adjusted centroids serve as our initialization for $\bmu_k$'s and thus, $\bmu$.

\subsection{Other computational settings} \label{sec:hyperparameters}

Our calibration of the prior hyperparameters follows some of the data-driven strategies proposed in \cite{chipman2010bart} and \cite{linero2018bayesian}.
We first apply min-max transformation on the response as, $(\by-\min(\by)-0.5(\max(\by)-\min(\by)))/(\max(\by)-\min(\by))$ which makes it bounded in $[-0.5, 0.5]$. After fitting the model, we apply an inverse transformation on the estimated $f_g$'s. The covariates $\bx$, are similarly normalized to the range  %$(\bx-\min(\bx))/(\max(\bx)-\min(\bx))$ 
within $[0,1]$ for continuous and ordinal variables. In the case of nominal categories with $C$ levels, we form $C-1$ binary dummy variables that do not need any further scaling.  
Here, we discuss our procedure for specifying the prior variance $\sigma^2_d$ of $\theta_k$ when there are $K_1$ bases. Subsequently, we set a suitable value for $K_1$ based on $\gamma_{k,g}$'s.
Under the RBF-net model, the expectation of a response given the covariate $\bx$ with $K_1$ bases is $\sum_{k=1}^{K_1}\theta_{k}\exp\{-(\bx-\bmu_k)^T(\bx-\bmu_k)/b_k^2\}$. 
%The number of bases for each treatment group depends on the indicator matrix $\bGamma$.
For a random variable distributed as $\Normal(\mu,\sigma^2)$, most of its mass falls within $\mu\pm 1.96\sigma$. Furthermore, the maximum value of $\exp\{-(\bx_i-\bmu_k)^T(\bx_i-\bmu_k)/b_k^2\}$ is 1. Thus, the value of $\sum_{k=1}^{K_1}\theta_{k}\exp\{-(\bx-\bmu_k)^T(\bx-\bmu_k)/b_k^2\}$ will be within $0\pm 1.96\sigma_d\sqrt{K_1}$ if $\theta_k\sim\Normal(0,\sigma^2_d)$. Our min-max normalized data satisfies $|y_{\min}|=|y_{\max}|=0.5$. Thus, we may set $0.5=1.96\sigma_d\sqrt{K_1}$. Hence, approximately, $\sigma_d=1/(4\sqrt{K_1})=1/(\sqrt{16K_1})$. %To incorporate this, we let $\sigma_d\sim  C^{+}(0,A_d)$ and set $A_d=1/(\sqrt{16K_1})$.

Now, in our case, the numbers of active bases are different for different $f_g(\cdot)$'s. For group $g$, we have $\sum_{k=1}^K\gamma_{k,g}$ bases. We, thus, consider to initialize $K_1$ as $K_1=\max_g \sum_{k=1}^K\gamma_{k,g}$, based on our initialization of $\bGamma$. %In the above expression ($\max_g \sum_{k=1}^K\gamma_{k,g}$) for setting $K_1$, 
One may also consider mean or median over $g$ too, as we do not necessarily need $K_1$ to be an integer here. However, since this value may be inappropriate as parameters are updated, we recalculate $K_1$ during the burn-in phase of our MCM chain. %Afterwards, we set $K_1$ as $K$ for every 200 iterations.

For parameter initialization, $\btheta$ is set as the least squares estimate. We initialize the probabilities $p_g = 1/2$ for all $g$, allowing an equal likelihood for all $k\in \I_{g}$, as described in Section \ref{sec:specifyK}, to be included. The bandwidth parameters $b_k$'s are also pre-specified. For simplicity, we let $b_k=b$ for all $k$. Then, we first initialize $\bmu$ as discussed in Section~\ref{sec:specifyK} and set: $b = \frac{\sqrt{2}}{k(k-1)} \sum_{i=1}^{k-1} \sum_{j=i+1}^{k} \|\bmu_{i} - \bmu_{j}\|_2$, where $\|\cdot\|_2$ represents the Euclidean norm and $\bmu_{j}$ is $j$-$th$ column of $\bmu$. Some of the prior hyperparameters are set as $\sigma_{\mu}=1, c=d=1$.

\subsection{Inference for CATE using linear projections and thresholding} \label{sec:blp}

Following \cite{semenova2021debiased, cui2023estimating}, we adopt a linear projection-based approach for the inference on $\tau(\cdot)$ to circumvent the difficulties in non-parametric inferences and provide a meaningful interpretation and summary of treatment heterogeneity, as argued in \cite{semenova2021debiased}.%focusing on a low-dimensional summary of $\tau(\cdot)$, such as projections, certain difficulties in non-parametric inferences on $\tau(\bx)$ itself can be circumvented. 
 The best linear projection (BLP) of $\tau(\cdot)$ given a set of covariates $\bx_i$ is defined as $$\left\{\beta_0^*, \bbeta^*\right\}=\operatorname{argmin}_{\beta_0, \bbeta} \mathbb{E}\left[\left(\tau\left(\bx_i\right)-\beta_0-\bx_i \bbeta\right)^2\right],$$
%\begin{equation}\left\{\beta_0^*, \bbeta^*\right\}=\operatorname{argmin}_{\beta_0, \bbeta} \mathbb{E}\left[\left(\tau\left(\bx_i\right)-\beta_0-\bx_i \bbeta\right)^2\right],\end{equation} 
where  $\beta_0^*$ and $\bbeta^*$ are the intercept and the coefficient estimates our covariates, respectively.
One can estimate the BLP parameters by regressing the CATE estimates $\hat{\tau}\left(\bx_i\right)$ on $\bx_i$.

We modify this approach for our Bayesian setting with three treatments in our numerical experiment and MIMIC data application. For example, we may consider ${\tau}_{21}$ and ${\tau}_{31}$, comparing treatments 2 and 3 with treatment 1, respectively. Then, for the $b$-$th$ posterior sample, we obtain ${\tau}_{21, b}\left(\bx_i\right)$ and ${\tau}_{31, b}\left(\bx_i\right)$, based on the posterior samples of $f_{g}$'s for the $i$-$th$ patient. We next separately regress ${\btau}_{21, b}\left(\bx\right)$ and ${\btau}_{31,b}\left(\bx\right)$ on $\bx$, providing the estimated BLP coefficients for each predictor.

Furthermore, we propose thresholding the BLP coefficients and computing the credible intervals.
We generate $B$ posterior samples and then compute the BLP coefficient $\beta^{(b)}_{p}$ for each predictor $p$ and posterior sample $b$. Finally, for a given predictor $p$ and threshold $t$, we compute $s_{p,t}=\frac{1}{B}\sum_{b=1}^B\bone\{|\beta^{(b)}_p|>t\}$, the proportion of coefficients falling outside of the interval $(-t,t)$ across all the posterior samples. A larger value of $s_{p,t}$ would indicate the higher importance of $p$-$th$ predictor under a given level of threshold $t$. It is easy to see that as $t$ increases, $s_{p,t}$ will decrease. A slower decay in $s_{p,t}$ with increasing $t$ indicates greater importance, providing us a measure for comparing importances of different predictors. 

It is important to note that the BLPs are not posterior samples of any model parameter.
They are estimated based on a loss minimization and can be expressed as a fixed function of the posterior samples. Specifically, the function here is a linear function $h(\bz)=\bA\bz$, where $\bA=\bX(\bX^T\bX)^{-1}\bX^T$ and $\bX$ is the 
$n\times p$ design matrix with $n$ observations and $p$ predictors. For $b$-$th$ posterior sample and $(g,g')$ pair of treatments, we set $\bz=\{\tau_{gg',b}(\bx_i)\}_{i=1}^n$ and compute $h(\bz)$ to get the BLPs. %We have now provided these details and hope it will offer more clarity. 
The proposed BLP-based approach works surprisingly well in Section~\ref{sec:sec4.2}. However, it should be used with caution. 

\section{Simulation} \label{sec:simu}

In this section, we assess the effectiveness of our proposed RBF method by estimating model parameters using the training data and calculating predicted mean square errors on the test set. The mean square error (MSE) for treatment-groups pair $(g,g')$ is defined as $\frac{1}{\left|T_r\right|} \sum_{i \in T_r}\left(\tau_{gg'}(\bx_i)-\hat{\tau}_{gg'}(\bx_i)\right)^2$, where $T_r$ represents the test set for the $r$-$th$ replication. We then compute the median MSE over $r$ for each method. Here, $\hat{\tau}_{gg'}(\bx_i)$ represents the posterior mean for the Bayesian implementations and the point estimate for the frequentist implementations of $\tau_{gg'}(\bx_i)$, respectively.
We conduct two simulations with 50 replications.

In each case, we compare our estimates with a multi-arm causal forest from the {\tt grf} \citep{grfR} package-based solution. For CATE estimation \citep{grfR}, {\tt multi\_arm\_causal\_forest} and {\tt predict.multi\_arm\_causal\_forest} functions are applied on the training and testing set, respectively. 
To the best of our knowledge, there is no other package that supports CATE estimation for more than two treatments for out-of-sample predictors. Additionally, we compare these results with other methods using T-learning and S-learning.
We employ T-learning by fitting treatment-specific functions. 
For S-learning with three treatments, we model the outcome $y_i = f(\bx_i, z_{2i}, z_{3i})+\epsilon_i$, where $z_{2i}=1$ if $i$-$th$ subject is treated under the second treatment and zero otherwise; $z_{3i}$ is defined similarly for the third treatment. Consequently, $z_{2i}$ and $z_{3i}$ are both zero if $i$-$th$ subject is in the first treatment group. 

Our candidate models for both approaches are relevance vector machine (RVM), support vector regression (SVR), ordinary least square (OLS), and BART, using the corresponding R packages {\tt e1071R} \citep{e1071R}, {\tt kernlab} \citep{kernlab}, {\tt BART}\citep{BARTR}, and few other functions from the base packages of R.

The outcome $\by$ is generated by $y_i$ from $\Normal\left(f_g(\bx_i), 1\right)$, where three treatment-outcome functions $f_1(\cdot), f_2(\cdot),f_3(\cdot)$ are 

\begin{align*}
f_1(\bx) &=f_{1}^0(\bx) +  f_2^0(\bx), \\
f_2(\bx) &=(f_{1}^0(\bx) + f_2^0(\bx))/2 + f_3^0(\bx)/3, \\ 
f_3(\bx) &= (f_{2}^0(\bx) + f_3^0(\bx))/2 + f_1^0(\bx)/3,
\end{align*} 
where

\begin{align*}
f_{1}^0(\bx) &=10 \sin \left(\pi x_1 x_2\right)+20\left(x_3-0.5\right)^2+\left(10 x_4+5 x_5\right), \\
f_2^0(\bx) &=(5 x_2)/(1+x_1^2)+ 5\sin(x_3 x_4) + x_5, \\
f_3^0(\bx) &= 0.1\exp(4x_1) + 4/(1 + \exp(-20(x_2 - 0.5))) + 3x_3^2 + 2x_4 + x_5.
\end{align*}

\subsection{Setting 1} \label{sec:sec4.1}

Under this setting, we consider two sample sizes 180 and 360 with three equal treatment groups. To assess the performance of our model, each dataset is partitioned into training and testing sets with a 67/33 split. 

The results, as detailed in Table \ref{sim4.1tab}, demonstrate the efficacy of our proposed method. Notably, our approach consistently outperforms both S-learning and T-learning-type models, where treatment-specific separate outcome functions are fitted. This is particularly evident in the scenario with the smaller sample size.  Compared to the multi-arm causal forest, our method demonstrates superior performance in the smaller sample size case, while results are comparable for larger sample sizes. This indicates that our model is particularly effective with limited data, an important advantage for practical applications where large datasets may be scarce.

\begin{table}[ht]
\centering
\caption{Comparing estimation errors for the two treatment contrasts $\tau_{21}$ and $\tau_{31}$ across different methods, based on 50 replications and two choices for the total sample size under Simulation setting 1. The errors for the S-learning and T-learning type models are separated from the two multi-treatment cases with a horizontal line. The reported errors are the medians over all replications, with the corresponding ranges in parentheses.}
\begin{tabular}{ccccc}
        \hline 
        & \multicolumn{2}{c|}{$n = 180$} & \multicolumn{2}{c}{$n = 360$} \\
        \hline
        &  \multicolumn{1}{c|}{$\hat{\btau}_{21}$ MSE} &  \multicolumn{1}{c|}{$\hat{\btau}_{31}$ MSE} &  \multicolumn{1}{c|}{$\hat{\btau}_{21}$ MSE} & $\hat{\btau}_{31}$ MSE \\ 
        \hline
        Shared-neuron RBF & 2.99 & 4.84 & 3.45 & 4.51 \\ 
        & (2.36, 4.49) & (4.19, 5.94) & (2.89, 4.85) & (3.99, 5.41) \\ 
        Causal Random Forest & 5.32 & 9.9 & 2.54 & 5.41 \\
        & (5.01, 6.34) & (9.34, 10.61) & (2.18, 2.88) & (4.73, 6.27) \\      
        \hline
        RVM (S) & 5.34 & 10.3 & 6.95 & 7.22 \\ 
        & (3.28, 35.01) & (5.88, 13.37) & (3.2, 10.81) & (4.77, 10.49) \\          
        SVR (S) & 5.11 & 7.34 & 6.29 & 7.75 \\ 
        & (3.51, 10.13) & (4.93, 10.41) & (4.5, 7.59) & (6.61, 10.76) \\         
        OLS (S) & 7.3 & 10.25 & 10.24 & 11.54 \\
        & (6.36, 8.76) & (10.13, 10.66) & (9, 11.57) & (10.74, 12.46) \\        
        BART20 (S) & 4.43 & 9.11 & 8.66 & 8.86 \\ 
        & (2.82, 8.88) & (6.13, 12.74) & (5.58, 11.91) & (4.82, 12.16) \\       
        BART50 (S) & 3.26 & 8.07 & 5.02 & 5.56 \\ 
        & (2.21, 4.5) & (7.27, 9.33) & (3.66, 7.84) & (3.85, 7.77) \\         
        BART200 (S) & 3.63 & 8.61 & 2.07 & 4.69 \\ 
        & (3.03, 4.15) & (8.22, 9.11) & (1.54, 2.58) & (3.94, 5.39) \\       
        \hline
        RVM (T) & 14.03 & 10.96 & 8.71 & 9.75 \\ 
        & (11.05, 19.75) & (9.3, 13.62) & (7.81, 12.1) & (7.04, 12.78) \\       
        SVR (T) & 5.68 & 7.11 & 8.87 & 8.98 \\ 
        & (3.91, 9.82) & (5.23, 9.48) & (7.01, 11.22) & (6.79, 11.92) \\        
        OLS (T) & 8.8 & 9.01 & 4.97 & 5.93 \\ 
        & (6.16, 12.45) & (6.48, 12.27) & (3.98, 6.61) & (4.74, 8.19) \\        
        BART20 (T) & 14.55 & 8.8 & 11.14 & 9.47 \\ 
        & (11.28, 17.96) & (6.3, 12.24) & (8.8, 14.27) & (6.25, 11.67) \\         
        BART50 (T) & 14.59 & 8.19 & 10.16 & 8.38 \\ 
        & (11.63, 17.92) & (5.86, 11.03) & (8.41, 12.59) & (5.93, 10.19) \\         
        BART200 (T) & 15.07 & 8 & 10.85 & 8.48 \\ 
        & (11.66, 18.24) & (6.1, 10.99) & (9.06, 13.19) & (6.33, 10.09) \\ 
\end{tabular}
\label{sim4.1tab}
\end{table}

\subsection{Setting 2}\label{sec:sec4.2}

In Setting 1, the assignment of the treatment groups and the generation of the predictors are performed independently. We now use the MIMIC-III database to evaluate our model with unknown, real-world complexities in predictor-treatment dependencies. 
%based on outcomes of length of stay in the ICU and SOFA score for septic patients at their 12th hour of their admission. 
From the dataset, we constructed three treatment groups based on the usage of only mechanical ventilation (group 1), only vasopressor (group 2), and both (group 3), as detailed in Section~\ref{sec:real}. Our final dataset in our real data analysis comprises of 172 subjects and 15 predictors but for this simulation, we use a subset of predictors. Employing the marginal screening approach from \cite{xue2017robust}, we pick the top five most important predictors for each of the two outcomes: the length of stay in the Intensive Care Unit (LOS-ICU) and Sequential Organ Failure Assessment (SOFA) scores.
%For the selection of predictors, we adopt the marginal screening approach as introduced by \cite{xue2017robust}. The initial step involves the application of a nonparanormal (npn) transformation to both the set of predictors and outcomes. This transformation is executed using the {\tt huge.npn} function within the R package {\tt huge} \citep{zhao2012huge}.  Subsequently, the Henze-Zirkler (HZ) test is conducted on each pairwise combination of outcomes and predictors. The {\tt HZ.test} from the R package {\tt mvnTest} is used for this test \citep{henze1990class}. 
After taking the union, we find the following set of seven key predictors: age, weight, bicarbonate level, sodium level, systolic blood pressure (SBP), potassium level, and mean arterial pressure (MAP). 

We then generate $\by$ in the same manner as in the preceding section based on the first five predictors% age, weight, bicarbonate level, sodium level, and SBP
(excluding potassium and MAP), but fit the model with all seven to evaluate their importance subsequently.  We want to mention here that the marginal screening procedure is not essential for this simulation but is used systematically to ensure we include the most important predictors associated with the primary outcomes in Section~\ref{sec:real}. This approach enables complex associations between predictors and treatment assignments, as they are also borrowed from MIMIC data.

%The true outcome functions are identical to those in Section~\ref{sec:sec4.1}. 
Table \ref{sim4.2tab} presents an MSE comparison of the CATE estimates of our method versus other competing methods. As in the smaller sample size case of Section \ref{sec:sec4.1},  our model consistently outperforms the multi-arm causal forest, and among S and T-learning methods, our method excels except for the marginally better performance by support vector regression under T-learning.

%Furthermore, we apply the projection-based inference method from Section~\ref{sec:blp}. Specifically, 
Since the outcome $\by$ depends on only the first five predictors, we turn to our thresholding-based credible interval comparison approach from Section~\ref{sec:blp} to analyze the $s_{p,t}$-values for different predictors ($p$) and threshold ($t$), ranging from 0.1 to 2.0. Table \ref{sim4.2threshold} shows the $s_{t,p}$-values based on 50 replications for $\btau_{23}$ and $\btau_{13}$, respectively. 
As expected, the decrease in $s_{p,t}$ with $t$ is the sharpest for potassium and MAP, which aligns with their exclusion from the data generation process. 
The $s_{p,t}$-values thus provide useful insights into the relationship between the outcome and the predictors.

Here, we show that even flexible methods effective in treatment-effect estimation may underperform if the response curves share common characteristics. For example, while the RVM model with a Gaussian kernel resembles our RBF model, both T and S learning implementations of RVM are less efficient than the shared-kernel model. Since treatments often share commonalities, our proposed method with shared bases outperforms alternatives lacking shared structural features.
%Although it is not an optimal approach for identifying important predictors, we still obtain useful insights into the relationship between the outcome and the predictors.

\begin{table}[ht]
\centering
\caption{Comparing estimation errors for the two treatment contrasts $\tau_{23}$ and $\tau_{13}$ across different methods, based on 50 replications and two choices for the total sample size under Simulation setting 2. The errors for the S-learning and T-learning type models are separated from the two multi-treatment cases with a horizontal line.  The reported errors are the medians over all replications, with the corresponding ranges in parentheses.}
\begin{tabular}{ccc}
    \hline
    &  \multicolumn{1}{c|}{$\hat{\btau}_{23}$ MSE} & $\hat{\btau}_{13}$ MSE \\ 
    \hline
  Shared-neuron RBF & 2.97 & 4.28 \\ 
  & (2.37, 4.07) & (2.47, 6.73) \\ 
  Causal Random Forest & 4.67 & 10.28 \\ 
  & (4.6, 5.35) & (6.93, 14.03) \\ 
  \hline
  RVM (S) & 4.53 & 10.35 \\ 
  & (3.65, 53.73) & (6.48, 47.12) \\   
  SVR (S) & 11.21 & 13.89 \\ 
  & (10.37, 13.21) & (10.75, 18.9) \\ 
  OLS (S) & 4.82 & 8.06 \\
  & (4.79, 5.22) & (5.57, 10.6) \\ 
  BART20 (S) & 2.97 & 9.1 \\ 
  & (2.3, 3.61) & (5.2, 13.24) \\ 
  BART50 (S). & 3.29 & 11.67 \\ 
  & (2.56, 4.03) & (7.35, 16.69) \\ 
  BART200 (S) & 3.85 & 13.08 \\ 
  & (3.31, 5.02) & (8.49, 17.6) \\ 
  \hline
  RVM (T) & 12.99 & 9.91 \\ 
  & (8.52, 20.08) & (7.25, 14.64) \\ 
  SVR (T) & 2.63 & 3.87 \\ 
  & (1.7, 4.62) & (2.3, 6.36) \\ 
  OLS (T) & 6.51 & 9.57 \\
  & (4.43, 9.76) & (6.09, 13.39) \\ 
  BART20 (T) & 11.1 & 12.47 \\ 
  & (5.94, 17.15) & (9, 19.57) \\ 
  BART50 (T) & 10.46 & 12.15 \\ 
  & (6.51, 16.73) & (8.77, 15.57) \\ 
  BART200 (T) & 9.7 & 11.72 \\ 
  & (6.75, 15.38) & (8.56, 15.45) \\ 
\end{tabular}
\label{sim4.2tab}
\end{table}

\begin{table}[ht]
\footnotesize
\centering
\caption{Comparison of thresholding-based $s_{p,t}$-values for $\hat{\btau}_{23}$ and $\hat{\btau}_{13}$ for different choices of thresholds ($t$), and predictors. The thresholding levels are given in the first column. The generated outcomes are independent of the last two predictors. The exclusion proportions are based on 50 replicated datasets.}
\begin{tabular}{ccccccccc}
  \hline
 \multicolumn{1}{c|}{$\hat{\btau}_{23}$} &  \multicolumn{1}{c|}{Intercept} &  \multicolumn{1}{c|}{SBP} &  \multicolumn{1}{c|}{Bicarbonate} &  \multicolumn{1}{c|}{Sodium} &  \multicolumn{1}{c|}{Age} &  \multicolumn{1}{c|}{Weight} &  \multicolumn{1}{c|}{Potassium} &  \multicolumn{1}{c|}{MAP} \\
\hline
 0.1 & 1.00 & 1.00 & 1.00 & 1.00 & 1.00 & 1.00 & 0.95 & 1.00 \\ 
  0.2 & 1.00 & 1.00 & 1.00 & 1.00 & 1.00 & 1.00 & 0.89 & 1.00 \\ 
  0.3 & 1.00 & 1.00 & 1.00 & 1.00 & 1.00 & 1.00 & 0.84 & 1.00 \\ 
  0.4 & 1.00 & 1.00 & 1.00 & 0.99 & 1.00 & 1.00 & 0.79 & 1.00 \\ 
  0.5 & 1.00 & 1.00 & 1.00 & 0.99 & 1.00 & 1.00 & 0.74 & 0.99 \\ 
  0.6 & 1.00 & 0.99 & 1.00 & 0.99 & 1.00 & 1.00 & 0.69 & 0.99 \\ 
  0.7 & 1.00 & 0.99 & 1.00 & 0.99 & 1.00 & 1.00 & 0.64 & 0.99 \\ 
  0.8 & 1.00 & 0.99 & 1.00 & 0.99 & 1.00 & 1.00 & 0.59 & 0.98 \\ 
  0.9 & 1.00 & 0.99 & 1.00 & 0.98 & 1.00 & 1.00 & 0.55 & 0.97 \\ 
  1.0 & 1.00 & 0.99 & 1.00 & 0.98 & 1.00 & 1.00 & 0.50 & 0.96 \\ 
  1.1 & 1.00 & 0.99 & 1.00 & 0.97 & 1.00 & 1.00 & 0.46 & 0.95 \\ 
  1.2 & 1.00 & 0.98 & 1.00 & 0.97 & 1.00 & 1.00 & 0.42 & 0.93 \\ 
  1.3 & 1.00 & 0.98 & 1.00 & 0.96 & 1.00 & 0.99 & 0.38 & 0.91 \\ 
  1.4 & 1.00 & 0.98 & 1.00 & 0.96 & 1.00 & 0.99 & 0.34 & 0.88 \\ 
  1.5 & 1.00 & 0.98 & 1.00 & 0.95 & 1.00 & 0.99 & 0.30 & 0.84 \\ 
  1.6 & 1.00 & 0.97 & 1.00 & 0.94 & 1.00 & 0.99 & 0.27 & 0.80 \\ 
  1.7 & 1.00 & 0.97 & 1.00 & 0.93 & 1.00 & 0.99 & 0.24 & 0.76 \\ 
  1.8 & 1.00 & 0.97 & 1.00 & 0.92 & 1.00 & 0.99 & 0.21 & 0.70 \\ 
  1.9 & 0.99 & 0.96 & 1.00 & 0.91 & 1.00 & 0.98 & 0.19 & 0.65 \\ 
  2.0 & 0.99 & 0.96 & 1.00 & 0.89 & 1.00 & 0.98 & 0.16 & 0.59 \\ 
\hline
 \multicolumn{1}{c|}{$\hat{\btau}_{13}$} \\ 
  \hline
  0.1 & 1.00 & 1.00 & 1.00 & 1.00 & 1.00 & 1.00 & 0.95 & 1.00 \\ 
  0.2 & 1.00 & 1.00 & 1.00 & 1.00 & 1.00 & 1.00 & 0.90 & 1.00 \\ 
  0.3 & 1.00 & 1.00 & 1.00 & 1.00 & 1.00 & 1.00 & 0.84 & 1.00 \\ 
  0.4 & 1.00 & 1.00 & 1.00 & 1.00 & 1.00 & 1.00 & 0.79 & 0.99 \\ 
  0.5 & 1.00 & 1.00 & 1.00 & 0.99 & 1.00 & 1.00 & 0.74 & 0.99 \\ 
  0.6 & 1.00 & 1.00 & 1.00 & 0.99 & 1.00 & 1.00 & 0.69 & 0.98 \\ 
  0.7 & 1.00 & 1.00 & 1.00 & 0.99 & 1.00 & 1.00 & 0.64 & 0.98 \\ 
  0.8 & 1.00 & 0.99 & 1.00 & 0.99 & 1.00 & 1.00 & 0.59 & 0.97 \\ 
  0.9 & 1.00 & 0.99 & 1.00 & 0.98 & 1.00 & 1.00 & 0.54 & 0.96 \\ 
  1.0 & 1.00 & 0.99 & 1.00 & 0.98 & 1.00 & 1.00 & 0.49 & 0.94 \\ 
  1.1 & 1.00 & 0.99 & 1.00 & 0.98 & 1.00 & 1.00 & 0.45 & 0.93 \\ 
  1.2 & 1.00 & 0.99 & 1.00 & 0.97 & 1.00 & 1.00 & 0.41 & 0.90 \\ 
  1.3 & 1.00 & 0.99 & 1.00 & 0.97 & 1.00 & 1.00 & 0.37 & 0.87 \\ 
  1.4 & 1.00 & 0.99 & 1.00 & 0.96 & 1.00 & 1.00 & 0.33 & 0.84 \\ 
  1.5 & 1.00 & 0.99 & 1.00 & 0.95 & 1.00 & 0.99 & 0.29 & 0.79 \\ 
  1.6 & 1.00 & 0.98 & 1.00 & 0.94 & 1.00 & 0.99 & 0.26 & 0.75 \\ 
  1.7 & 1.00 & 0.98 & 1.00 & 0.93 & 1.00 & 0.99 & 0.23 & 0.69 \\ 
  1.8 & 1.00 & 0.98 & 1.00 & 0.92 & 1.00 & 0.99 & 0.20 & 0.64 \\ 
  1.9 & 1.00 & 0.98 & 1.00 & 0.91 & 1.00 & 0.99 & 0.17 & 0.58 \\ 
  2.0 & 1.00 & 0.97 & 1.00 & 0.89 & 1.00 & 0.98 & 0.15 & 0.52 \\ 
   \hline
\end{tabular}
\label{sim4.2threshold}
\end{table}

\section{MIMIC Data Analysis} \label{sec:real}

MIMIC-III is a large single-center database, containing deidentified, comprehensive clinical data of 53,423 adult ICU patient admissions from 2001 to 2012 at Beth Israel Deaconess Medical Center in Boston, Massachusetts \citep{johnson2016mimic}. The dataset offers a vast amount of patient information, encompassing demographics, laboratory results, medicine summaries, and diagnostic codes. This extensive data allows us to employ our RBF-net with shared neurons for septic patients.
With the most current revision, sepsis is defined following \cite{singer2016third} as a life-threatening condition resulting from dysregulated host responses to infection, often identified by a Sequential [Sepsis-related] Organ Failure Assessment (SOFA) score increase of 2 points or more. Its progression can lead to septic shock, marked by severe circulatory and metabolic dysfunction. 
Our analysis focuses on septic patients' data from the first 12 hours of their initial visit, excluding readmissions. We define sepsis based on diagnoses of ``Sepsis", ``Septic shock", or ``Severe sepsis" during their inaugural visit and include only those discharged to their ``home."

One challenge regarding MIMIC-III, and electronic health records in general, is the amount of missing data \citep{groenwold2020informative}. To handle missing data in MIMIC-III \citep{hou2020predicting}, %li2021prediction}, 
we exclude variables with more than 20\% missing data and apply a hybrid procedure:  first, the last observation carried forward (LOCF) using the {\tt na.locf} function from the R package {\tt zoo} \citep{zoo}, followed by multiple imputations for patients who do not have observations in the initial hours of their visit. We consider multiple imputations using the R package {\tt mice} \citep{mice}. 

The set of predictors considered in our analysis are age, weight, hemoglobin, white blood cells, blood urea nitrogen (BUN), potassium, sodium, bicarbonate, creatinine, platelet count, heart rate, systolic blood pressure (SBP), diastolic blood pressure (DBP), mean arterial pressure (MAP), and temperature. %Among these variables, the maximum percentage of missing is found to be 8.14\% for the `white blood cell' variable as shown in Table \ref{mimic_dem}. 
As discussed in Section \ref{sec:sec4.2}, our treatment groups are based on the usage of only mechanical ventilation (group 1), only vasopressor (group 2), and both (group 3). Tables \ref{mimic_dem} and \ref{mimic_dem2} are the demographics for our study population, with Table \ref{mimic_dem2} stratifying by treatment group. The demographics contain the mean and standard deviation of both the predictors and outcomes. 

\begin{table}[ht]
\centering
\caption{Overall Demographics for Selected Septic Patients in MIMIC-III Database. Predictor variables reported represent the average value after 12 hours of observation in hospital. Outcome variables reported represent the value at the 12th hour. We also report the amount of missing values in each group before applying LOCF and imputation to complete the data.}
\begin{tabular}{rrr}
  \hline 
        & \multicolumn{2}{c}{\textbf{Overall}} \\
  \hline
  \multicolumn{1}{c|}{\textbf{Variable}} &  \multicolumn{1}{c|}{Mean (SD)} &\% Missing\\ 
  \hline
    Age & 53.8 (17.78) & 0\% \\ 
    Bicarbonate & 20.04 (4.15) & 4.65\% \\ 
    BUN & 25.43 (18.68) & 4.65\% \\ 
    Creatinine & 1.58 (1.54) & 4.65\% \\ 
    DBP & 60.57 (8.67) & 0\% \\ 
    Hemoglobin & 10.86 (1.81) & 7.56\% \\ 
    Heart Rate & 93.26 (17.91) & 0\% \\ 
    Potassium & 4.04 (0.67) & 4.65\% \\ 
    Platelet Count & 191.86 (104.14) & 7.56\% \\ 
    MAP & 73.74 (9.15) & 0\% \\ 
    SBP & 107.58 (11.97) & 0\% \\ 
    Sodium & 138.86 (3.6) & 5.23\% \\ 
    Temperature & 37.06 (0.82) & 0\% \\ 
    Weight & 82.29 (21.67) & 2.91\% \\ 
    White Blood Cell & 14.9 (8.44) & 8.14\% \\ 
    \hline
    Length of Stay in ICU & 4.47 (4.01) & 0\% \\ 
    SOFA Score & 6.96 (2.82) & 0\% \\ 
   \hline
\end{tabular}

\label{mimic_dem}
\end{table}

\begin{table}[ht]
\footnotesize
\centering
\caption{Treatment group-specific summaries of the covariates for selected sets of septic patients in the MIMIC-III database are illustrated here. For all the predictor variables, the reported statistics are based on the first 12-hour averages under ICU. The statistics on the SOFA score are evaluated based on the value at the 12th hour in the ICU.} %We also report the percentage of missing values in each group in the raw data.}
\begin{tabular}{rrrr}
  \hline 
  & \multicolumn{1}{c|}{\textbf{Ventilator Only}} & \multicolumn{1}{c|}{\textbf{Vasopressor Only}} &  \multicolumn{1}{c}{\textbf{Both}}\\
    \hline
  \multicolumn{1}{c|}{\textbf{Variable}} & \multicolumn{1}{c|}{Mean (SD)} & \multicolumn{1}{c|}{Mean (SD)}  & Mean (SD) \\ 
  \hline
  Age & 53.88 (19.41) & 46.7 (16.18) & 57.49 (16.63) \\ 
  Bicarbonate & 18.96 (4.79) & 21.48 (4.91) & 19.85 (3.02) \\ 
  BUN & 28.96 (21.08) & 23.84 (22.50) & 24.13 (14.44) \\ 
  Creatinine & 1.68 (1.06) & 1.42 (1.56) & 1.62 (1.76) \\ 
  DBP & 59.78 (8.87) & 64.56 (9.72) & 59.01 (7.48) \\ 
  Hemoglobin & 11.4 (2.28) & 10.82 (1.73) & 10.58 (1.50) \\ 
  Heart Rate & 93.96 (18.48) & 99.72 (18.78) & 89.57 (16.27) \\ 
  Potassium & 4.23 (0.78) & 4.07 (0.62) & 3.88 (0.60) \\ 
  MAP & 73.27 (9.07) & 79.62 (11.58) & 71.07 (5.84) \\ 
  Sodium & 138.71 (4.02) & 139.21 (4.34) & 138.84 (3) \\ 
  Platelet Count & 191.44 (89.81) & 212.27 (128.32) & 181.59 (98.06) \\ 
  SBP & 105.43 (8.76) & 117.01 (16.85) & 104.17 (6.96) \\ 
  Temperature & 37.02 (0.82) & 37.19 (0.76) & 37.0 (0.76) \\ 
  White Blood Cell & 13.36 (7.41) & 15.06 (9.12) & 15.65 (8.72) \\ 
  Weight & 84.99 (26.85) & 86.15 (25.24) & 78.95 (15.51) \\ 
  \hline
  Length of Stay in ICU & 6.14 (3.91) & 6.43 (5.5) & 2.48 (1.39) \\ 
  SOFA Score & 8.98 (2.82) & 6.05 (2.8) & 6.28 (2.22) \\ 
   \hline
\end{tabular}
\label{mimic_dem2}
\end{table}

For our analysis, we consider two possible outcomes, the length of ICU stay and the 12-$th$ hour SOFA score. We initially fit the model for both of the two outcomes in their original scale and the log scale. However, after examining the distributions of the residuals, we only consider reporting our results for the LOS-ICU in log scale and SOFA scores with a square root transformation as their residuals uphold the normality assumption. To further confirm this, we perform the Shapiro-Wilks test using {\tt shapiro.test} in R. The residual density plots are provided in Figures 1 and 2
in the supplementary material.

For each choice of outcome, we obtain $s_{p,t}$-values along with the BLPs for each covariate along for both $\hat{\btau}_{23}$ and $\hat{\btau}_{13}$.
The $s_{p,t}$ values illustrated in Section~\ref{sec:blp} are presented in Table \ref{mimic_bs} for different predictors ($p$) and thresholds ($t$) up to 0.3. For LOS-ICU, heart rate, MAP, and SBP show the highest $s_{p,t}$-values for $\hat{\btau}_{23}$, whereas most predictors for $\hat{\btau}_{13}$ experience a significant drop-off, leaving heart rate as the only predictor with non-zero $s_{p,t}$-values at higher thresholds. When analyzing SOFA score, predictors such as bicarbonate, creatinine, MAP, platelet count, and SBP exhibit the largest $s_{p,t}$-values in both $\hat{\btau}_{23}$ and $\hat{\btau}_{13}$. Notably, platelet count has non-zero $s_{p,t}$-values at threshold 0.3 for both $\hat{\btau}_{23}$ and $\hat{\btau}_{13}$.
%At lower thresholds, SBP emerges as the most important variable for $\hat{\btau}_{23}$ in both outcomes. However, at higher thresholds, heart rate and platelet count become the most important variable for LOS-ICU and SOFA scores, respectively.
The BLP coefficients for these important predictors are presented in Table \ref{mimic_blp}. For LOS-ICU, heart rate and MAP both have negative BLP coefficients. In contrast, SBP has a positive BLP coefficient. For the square-root transformed SOFA score, the BLP coefficients of platelet count and SBP are both positive. whereas, MAP has a negative coefficient. Their implications are further discussed in the context of optimal treatments in Section \ref{sec:discussion}.

\begin{table}[ht]
\footnotesize
\centering
\caption{Comparison of exclusion proportions for thresholding-based $s_{p,t}$-values for $\hat{\btau}_{23}$ and $\hat{\btau}_{13}$ for log(LOS-ICU) and $\sqrt{\text{SOFA Score}}$ for different choices of thresholds ($t$), and predictors. The thresholding levels are given in the first row. Treatment groups are based on the usage of only mechanical ventilation (group 1), only vasopressor (group 2), and both (group 3).}
\begin{tabular}{rcccccccccccc}
\hline
 & \multicolumn{12}{c}{\textbf{$\log$(LOS-ICU)}}\\ 
 \hline
 & \multicolumn{6}{c|}{\textbf{$\hat{\btau}_{23}$}} & \multicolumn{6}{c}{\textbf{$\hat{\btau}_{13}$}}\\
 \hline
 \textbf{Variable} &  0 &  0.1 &  0.15 &  0.2 &  0.25 &  \multicolumn{1}{c|}{0.3} &  0 &  0.1 &  0.15 &  0.2 &  0.25 &  0.3 \\ 
  \hline
  Intercept & 1.00 & 0.94 & 0.92 & 0.89 & 0.85 & 0.82 & 0.99 & 0.46 & 0.34 & 0.23 & 0.16 & 0.12 \\ 
  Age & 1.00 & 0.59 & 0.41 & 0.23 & 0.11 & 0.04 & 0.99 & 0.04 & 0.00 & 0.00 & 0.00 & 0.00 \\ 
  Bicarbonate & 1.00 & 0.43 & 0.21 & 0.08 & 0.02 & 0.00 & 0.99 & 0.01 & 0.00 & 0.00 & 0.00 & 0.00 \\ 
  BUN & 1.00 & 0.44 & 0.20 & 0.05 & 0.01 & 0.00 & 0.99 & 0.00 & 0.00 & 0.00 & 0.00 & 0.00 \\ 
  Creatinine & 1.00 & 0.48 & 0.33 & 0.22 & 0.14 & 0.08 & 0.99 & 0.01 & 0.00 & 0.00 & 0.00 & 0.00 \\ 
  DBP & 1.00 & 0.57 & 0.41 & 0.24 & 0.12 & 0.03 & 0.99 & 0.04 & 0.00 & 0.00 & 0.00 & 0.00 \\ 
  Hemoglobin & 1.00 & 0.57 & 0.39 & 0.18 & 0.07 & 0.02 & 0.99 & 0.06 & 0.01 & 0.00 & 0.00 & 0.00 \\ 
  Heart Rate & 1.00 & 0.79 & 0.68 & 0.56 & 0.46 & 0.36 & 0.99 & 0.24 & 0.12 & 0.05 & 0.02 & 0.01 \\ 
  MAP & 1.00 & 0.87 & 0.76 & 0.60 & 0.45 & 0.27 & 0.99 & 0.03 & 0.00 & 0.00 & 0.00 & 0.00 \\ 
  Potassium & 1.00 & 0.50 & 0.23 & 0.07 & 0.02 & 0.00 & 0.99 & 0.00 & 0.00 & 0.00 & 0.00 & 0.00 \\ 
  Platelet Count & 1.00 & 0.57 & 0.39 & 0.21 & 0.09 & 0.02 & 0.99 & 0.00 & 0.00 & 0.00 & 0.00 & 0.00 \\ 
  SBP & 1.00 & 0.81 & 0.72 & 0.60 & 0.49 & 0.35 & 0.99 & 0.04 & 0.01 & 0.00 & 0.00 & 0.00 \\ 
  Sodium & 1.00 & 0.49 & 0.30 & 0.14 & 0.05 & 0.01 & 0.99 & 0.01 & 0.00 & 0.00 & 0.00 & 0.00 \\ 
  Temperature & 1.00 & 0.43 & 0.20 & 0.06 & 0.01 & 0.00 & 0.99 & 0.00 & 0.00 & 0.00 & 0.00 & 0.00 \\ 
  Weight & 1.00 & 0.52 & 0.30 & 0.13 & 0.03 & 0.00 & 0.99 & 0.00 & 0.00 & 0.00 & 0.00 & 0.00 \\ 
  White Blood Cell & 1.00 & 0.54 & 0.41 & 0.25 & 0.12 & 0.05 & 0.99 & 0.00 & 0.00 & 0.00 & 0.00 & 0.00 \\ 
  \hline
  & \multicolumn{12}{c}{$\sqrt{\textbf{SOFA Score}}$} \\ 
  \hline
  Intercept & 1.00 & 0.79 & 0.71 & 0.64 & 0.55 & 0.46 & 1.00 & 0.82 & 0.74 & 0.67 & 0.60 & 0.52 \\ 
  Age & 1.00 & 0.03 & 0.00 & 0.00 & 0.00 & 0.00 & 1.00 & 0.05 & 0.00 & 0.00 & 0.00 & 0.00 \\ 
  Bicarbonate & 1.00 & 0.36 & 0.15 & 0.05 & 0.02 & 0.00 & 1.00 & 0.32 & 0.12 & 0.04 & 0.01 & 0.00 \\ 
  BUN & 1.00 & 0.13 & 0.02 & 0.00 & 0.00 & 0.00 & 1.00 & 0.30 & 0.12 & 0.05 & 0.01 & 0.00 \\
  Creatinine & 1.00 & 0.29 & 0.12 & 0.05 & 0.01 & 0.00 & 1.00 & 0.51 & 0.35 & 0.22 & 0.12 & 0.07 \\ 
  DBP & 1.00 & 0.01 & 0.00 & 0.00 & 0.00 & 0.00 & 1.00 & 0.03 & 0.00 & 0.00 & 0.00 & 0.00 \\ 
  Hemoglobin & 1.00 & 0.03 & 0.00 & 0.00 & 0.00 & 0.00 & 1.00 & 0.04 & 0.00 & 0.00 & 0.00 & 0.00 \\ 
  Heart Rate & 1.00 & 0.02 & 0.00 & 0.00 & 0.00 & 0.00 & 1.00 & 0.06 & 0.00 & 0.00 & 0.00 & 0.00 \\ 
  MAP & 1.00 & 0.22 & 0.01 & 0.00 & 0.00 & 0.00 & 1.00 & 0.47 & 0.09 & 0.00 & 0.00 & 0.00 \\ 
  Potassium & 1.00 & 0.08 & 0.01 & 0.00 & 0.00 & 0.00 & 1.00 & 0.20 & 0.06 & 0.01 & 0.00 & 0.00 \\ 
  Platelet Count & 1.00 & 0.43 & 0.24 & 0.11 & 0.05 & 0.02 & 1.00 & 0.56 & 0.38 & 0.26 & 0.15 & 0.08 \\ 
  SBP & 1.00 & 0.63 & 0.17 & 0.01 & 0.00 & 0.00 & 1.00 & 0.75 & 0.33 & 0.05 & 0.00 & 0.00 \\ 
  Sodium & 1.00 & 0.00 & 0.00 & 0.00 & 0.00 & 0.00 & 1.00 & 0.00 & 0.00 & 0.00 & 0.00 & 0.00 \\ 
  Temperature & 1.00 & 0.01 & 0.00 & 0.00 & 0.00 & 0.00 & 1.00 & 0.01 & 0.00 & 0.00 & 0.00 & 0.00 \\ 
  Weight & 1.00 & 0.04 & 0.00 & 0.00 & 0.00 & 0.00 & 1.00 & 0.03 & 0.00 & 0.00 & 0.00 & 0.00 \\ 
  White Blood Cell & 1.00 & 0.02 & 0.00 & 0.00 & 0.00 & 0.00 & 1.00 & 0.02 & 0.00 & 0.00 & 0.00 & 0.00 \\ 
  \hline
\end{tabular}
\label{mimic_bs}
\end{table}

\begin{table}[ht]
\small
\centering
\caption{The BLP coefficients for predictors in MIMIC-III for $\btau_{23}(\bx)$ and $\btau_{13}(\bx)$ for log-transformed length of stay in the ICU and square-root transformed 12-th hour SOFA score. Treatment groups are based on the usage of only mechanical ventilation (group 1), only vasopressor (group 2), and both (group 3).}
\begin{tabular}{rcccc}
    \hline 
    & \multicolumn{2}{c|}{\textbf{$\hat{\btau}_{23}$}} & \multicolumn{2}{c}{\textbf{$\hat{\btau}_{13}$}}\\
  \hline
   \multicolumn{1}{c|}{\textbf{Variable}} &  \multicolumn{1}{c|}{\textbf{$\log$(LOS-ICU)}} &  \multicolumn{1}{c|}{$\sqrt{\textbf{SOFA Score}}$} &  \multicolumn{1}{c|}{\textbf{$\log$(LOS-ICU)}} & $\sqrt{\textbf{SOFA Score}}$\\ 
  \hline
Intercept & -0.77 & -0.52 & -0.04 & -0.17 \\ 
  Age & 0.08 & 0.02 & 0.00 & -0.02 \\ 
  Bicarbonate & -0.01 & -0.04 & 0.00 & -0.01 \\ 
  BUN & -0.00 & 0.00 & -0.00 & -0.02 \\ 
  Creatinine & -0.09 & -0.05 & -0.01 & -0.05 \\ 
  DBP & -0.07 & -0.03 & 0.00 & -0.00 \\ 
  Hemoglobin & 0.07 & 0.01 & 0.00 & -0.02 \\ 
  Heart Rate & -0.23 & -0.13 & -0.01 & -0.03 \\ 
  MAP & -0.23 & -0.15 & -0.01 & -0.05 \\ 
  Potassium & 0.03 & 0.02 & 0.00 & -0.01 \\ 
  Platelet Count & 0.09 & 0.03 & 0.00 & 0.03 \\ 
  SBP & 0.22 & 0.17 & 0.01 & 0.07 \\ 
  Sodium & -0.07 & -0.03 & 0.00 & 0.00 \\ 
  Temperature & 0.01 & -0.01 & -0.00 & -0.00 \\  
  Weight & 0.03 & 0.03 & 0.00 & 0.01 \\ 
  White Blood Cell & 0.11 & 0.04 & 0.00 & -0.01 \\ 
   \hline
\end{tabular}
\label{mimic_blp}
\end{table}

\section{Discussion} \label{sec:discussion}

Our findings indicate that the four physiological predictors MAP, SBP, heart rate, and platelet count may be the most important predictors for the two outcomes of interest. 
%Additionally, the BLP coefficients for these predictors guide us in identifying the optimal treatment plan among the three options. 
For both outcomes, a positive BLP coefficient indicates that administering both vasopressors and mechanical ventilation is preferred, while a negative coefficient suggests that either vasopressor or mechanical ventilation alone is preferable, depending on the specific context. 
For instance in LOS-ICU, heart rate and MAP both have negative BLP coefficients for $\widehat{\btau}_{23}$, indicating that patients with higher heart rate or MAP should receive vasopressors alone, as it is associated with shorter ICU stays. On the other hand, the positive coefficient of SBP suggests that a combination of vasopressor and mechanical ventilation is more beneficial for patients with higher SBP. For SOFA scores, platelet count and SBP both have positive BLP coefficients for $\widehat{\btau}_{23}$, suggesting that patients with higher values of these predictors would benefit from receiving both treatments, as this combination may lead to better outcomes in terms of SOFA scores. Conversely, the negative BLP coefficient for MAP in the SOFA score context indicates that patients with higher MAP should receive mechanical ventilation alone to reduce their SOFA scores. These findings align with previous studies that show how these predictors affect LOS-ICU and SOFA scores \citep{li2024persistent, khanna2023association, he2022platelet, zhang2022association}. 

For our inference, we proposed a Bayesian version of the BLP approach \citep{semenova2021debiased} with thresholding. However, it may be too simple to reflect the heterogeneity in the CATE. Recently, \cite{shin2024treatment} proposed a kernel-based approach for a similar problem but with an increased computational demand. Nevertheless, it will be useful to consider this approach in the future.

Looking ahead, we hope to potentially broaden our model's application. 
The shared basis construction can be generalized to other types of models as well. Like in tree-based models, one may allow sharing of trees.
Another direction will be to extend the proposed architecture to model other types of outcomes, such as right-censored survival data or binary or count-valued data. Another possible extension of the proposed model is to incorporate the covariate effect into $p_g$. They may help to further the scope and impact of our work, particularly in the field of personalized medicine.

\section*{Acknowledgement}
We thank the Editor, Associate Editor, and referee for the insightful comments that led to improvements in the content and presentation
of the paper.

\section*{Supplementary materials}

Supplementary materials some additional figures referenced in the main paper.

\section*{Conflict of interest}
None declared.

\section*{Data availability}
The Medical Information Mart for Intensive Care III (MIMIC-III) data are available through Physionet ({\url{mimic.physionet.org}}). The codes to run the model are available at {\url{https://github.com/pshuwei/RBF}} along with specific implementation codes for each simulation.

\newpage
\clearpage

\bibliographystyle{bibstyle}
\bibliography{main,BayesianCI_review}

\end{document}

\title{Supplementary Materials for `Individualized Multi-Treatment Response Curves Estimation using RBF-net with Shared Neurons'}

\author{PETER CHANG, ARKAPRAVA ROY\\
% Author addresses
\textit{Department of Biostatistics,
University of Florida,
%College of Public Health and Health Professions,
%Gainesville, Florida,
United States}
\\
% E-mail address for correspondence
{pchang27@ufl.edu, arkaprava.roy@ufl.edu}
}

% Running headers of paper:
\markboth%
% First field is the short list of authors
{CHANG, ROY}
% Second field is the short title of the paper
{Multi-treatment RBF-net: Supplementary}

\maketitle

\begin{figure}[!p]
\centering
\caption{Distribution of the average residuals $\frac{1}{S} \sum_s^S y_{i,s} - \hat{f}_{g,s}(\bx_i)$ obtained fitting the model, using the log-transformed length of stay in the ICU as the outcome. We generate 1000 MCMC samples from an original pool of 15000 samples, incorporating a burn-in phase of 5000 samples and thinning. These samples are then used to perform individual regressions to obtain the residuals for the $s$-th regression.}
\includegraphics[width=15cm]{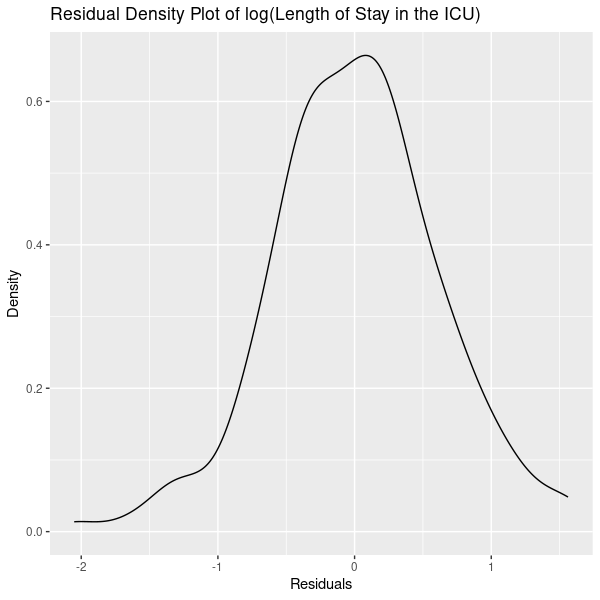}
\label{fig:logicu_res}
\end{figure}

\begin{figure}[!p]
\centering
\caption{Distribution of the average residuals $\frac{1}{S} \sum_s^S y_{i,s} - \hat{f}_{g,s}(\bx_i)$ obtained fitting the model, using the square-root transformed 12-th hour SOFA scores as the outcome. We generate 1000 MCMC samples from an original pool of 15000 samples, incorporating a burn-in phase of 5000 samples and thinning. These samples are then used to perform individual regressions to obtain the residuals for the $s$-th regression.}
\includegraphics[width=15cm]{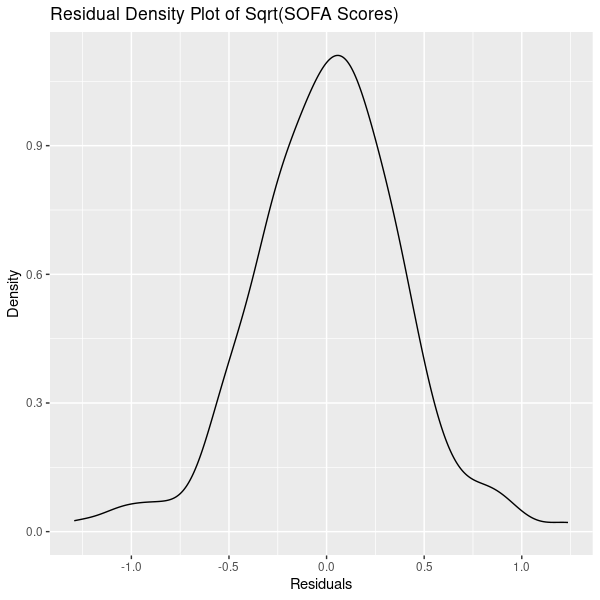}
\label{fig:sqrtsofa_res}
\end{figure}